\documentclass[11pt]{4ssmmp}

\usepackage{latexsym}

\newcommand{\beq}{\begin{equation}}
\newcommand{\eeq}{\end{equation}}
\newcommand\be{\begin{equation} }
\newcommand\bea{\begin{eqnarray}}
\newcommand\ee{\end{equation}}
\newcommand\eea{\end{eqnarray}}

\usepackage{epsfig}

\usepackage{fancyhdr}    
\usepackage{graphicx}
\usepackage{rotating}

\begin{document}
 \baselineskip=11pt

\title{Enveloping algebra Noncommutative SM: 
Renormalisability and High Energy Physics Phenomenology \hspace{.25mm}
\thanks{
Invited talk given at the 5th Mathematical Physics Meeting, 
Belgrad, Serbia, July 6-17, 2008.
This work is supported by the project
098-0982930-2900 of the Croatian Ministry of Science Education and Sports.
The author also acknowledge support from EU under the MRTN-CT-2006-035505
network programme.
}}
\author{\bf{Josip Trampeti\' c}\hspace{.25mm}\thanks{\,e-mail address: josipt@rex.irb.hr}
\\ \normalsize{Rudjer Bo\v skovi\' c Institute, Zagreb, Croatia} \vspace{2mm} \\ 
}

\date{}

\maketitle

\begin{abstract}
In this talk we discuss enveloping algebra based noncommutative gauge field theory, 
constructed at the first order in noncommutative parameter $\theta$,
as an effective, anomaly free theory, with one-loop renormalizable gauge sector. 
Limits on the scale of noncommutativity parameter 
$\Lambda_{\rm NC}$, via related phenomenology and associated 
experiments, are analyzed and a
firm bound to the scale of the noncommutativity is set around few TeV's.
\end{abstract}
\clearpage

 The Standard Model (SM) of particle physics and the theory of gravity
describe very well, as far as we know today, all physical phenomena 
from cosmological processes to the properties of subnuclear structures. 
Nevertheless, at extreme energy and/or
very short distances -at the Planck scale- this theories fail to be compatible, 
which motivates the study of
modified or alternative space-time structures
that could help to solve the above mentioned
difficulties or at least shed some light on them. 
These modified space-time structures arise in such frameworks as
the quantized coordinates  in string theory
or in deformation quantization.
The idea of noncommutative (NC) space-time, which can be realised in both of the above settings, 
has recently found more and more interest. In this paper we deal with noncommutative theories 
defined by means of the enveloping algebra approach, which allows to define gauge theories with 
arbitrary gauge groups, in particular that of the Standard Model. 
The research on these theories so far has successfully dealt 
with both theoretical and phenomenological aspects, 
which allow the confrontation of the theory with experiments.

One of the first example where noncommutativity (NC) was introduced is well known Heisenberg algebra.
Motivations to construct models on noncommutative space-time are comming from:  String Theory,
Quantum Gravity, Lorentz invariance breaking, and by its own right.
The star product ($\star$) definition is as usual. The $\star$-commutator  
and Moyal-Weyl $\star$-product of two functions are: 
\begin{eqnarray}   
[x^\mu\stackrel{\star}{,}x^\nu]  &=& x^\mu \star x^\nu - x^\nu \star x^\mu 
=ih\theta^{\mu\nu},
\;\;\;
\label{1}\\
(f \star g) (x) &=&  e^{-\frac{i}{2} \theta^{\mu\nu} 
\frac{\partial}{\partial{x^{\mu}}} \frac{\partial}{\partial{y^{\nu}}}}
f(x)g(y)|_{y \rightarrow x}\,.
\label{2}
\end{eqnarray} 
Here $\theta$ is constant, antisymmetric and real ${4\times4}$ matrix;  
${h=1/\Lambda^2_{\rm NC}}$ is noncommutative deformation parameter.
Symmetry in our model \cite{Wess}, using Seiberg-Witten map (SW) \cite{Seiberg:1999vs} 
is extended to enveloping algebra \cite{Wess,Calmet:2001na}.
Any enveloping algebra based model is essentialy double expansion
in power series in $\theta$ \cite{Wess,Calmet:2001na,Blazenka,Aschieri:2002mc,Goran}. 
In principle SW map express noncommutative functionals 
(parameters and functions of fields) 
spanned on the noncommutative space as a 
local functionals spanned on commutative space. 

To obtain the action we first do the Seiberg-Witten expansion of 
NC fields in terms of commutative ones and second we 
expand the $\star$-product. This procedure generets tower of new vertices, 
however it is valid for
any gauge group and arbitrary matter representation.
Also there is no charge quantization problem and no UV/IR mixing
\cite{UV/IR}.
Unitarity is satisfied for $\theta^{0i}=0$ and $\theta^{ij}\not=0$ 
\cite{Seiberg:2000gc,Gomis:2000zz}; however
careful canonical quantization produces always unitary theory. 
By covariant generalization of the condition $\theta^{0i}=0$ to: 
 \begin{eqnarray} 
\theta_{\mu\nu}\theta^{\mu\nu} &=&-\theta^2
= \frac{2}{\Lambda_{\rm NC}^4}
 \left (\vec{B}_{\theta}^2 - \vec{E}_{\theta}^2 \right ) >0\;, 
 \label{3}
\end{eqnarray}
which is known as {\it perturbative unitarity condition} \cite{Carroll:2001ws},
there is no difficulties with unitarity in NC gauge theories.
Finally covariant noncommutative Higgs and Yukawa couplings were constructed in \cite{Blazenka}.

There are two essential points in which NC gauge field theory (NCGFT) differ from
standard model (SM) gauge theories. 
The breakdown of Lorentz invariance with
respect to a fixed nonzero background field $\theta^{\mu\nu}$
(which fixes preferred directions) 
and the appearance of new interactions 
and the modification of standard ones. 
For example, triple--neutral--gauge boson, 
two fermion--two gauge bosons, 
direct photon-neutrino couplings, etc.
Both properties have a common origin and appear in a number of phenomena
at very high energies and/or very short distances.

In this article we consider $\theta$-expanded theories, constructed as an effective, 
anomaly free \cite{Brandt:2003fx} and one-loop renormalizable NCGFT 
\cite{Bichl:2001cq,Buric:2006wm,Latas:2007eu,Martin:2006gw},
at the first order in noncommutative parameter $\theta$. 
Finally we discuss related phenomenology and determine 
the scale of noncommutativity $\Lambda_{\rm NC}$, \cite{Buric:2007qx,Josip,Melic:2005hb}.\\


{\it Properties of $\theta$-expanded noncommutative gauge field theory satisfie}:

{\it Covariant coordinates} 
$ 
{\hat x}^\mu = x^\mu + h\theta^{\mu\nu} {\widehat A}_\nu 
$
were in noncommutative theory introduced in analogy to covariant derivatives in ordinary theory.

{\it Noncommutative gauge transformation}, i.e.
consider infinitesimal noncommutative local gauge 
transformation $\hat\delta$
of a fundamental matter field that carries a representation $\rho_\Psi$,
which is in Abelian case fixed by the hypercharge,
$
\hat\delta \widehat\Psi = i \rho_{\Psi}(\widehat\Lambda) \star \widehat\Psi \;.
$

{\it Locality of the theory}, i.e.
star-product of two ordinary functions $f(x)$ and $g(x)$,
determined by a Poisson tensor $\theta^{\mu\nu}$ and written in the form of expansion, 
is local function of 
$f$ and $g$ with finite number of derivatives at each order in $\theta$.

{\it Gauge equivalence for the theory}, i.e. 	
ordinary gauge
transformations $\delta A_\mu = \partial_\mu \Lambda + i[\Lambda,A_\mu]$
and $ \delta \Psi = i \Lambda\cdot \Psi$ 
induce noncommutative gauge transformations of the NC gauge and fermion fields 
$\widehat A$, $\widehat \Psi$ with NC gauge parameter $\widehat \Lambda$: 
$\delta \widehat A_\mu = \hat\delta \widehat A_\mu$ and
$\delta \widehat \Psi = \hat\delta \widehat \Psi$.

{\it Consistency condition} require that any pair of noncommutative 
gauge parameters 
$\widehat\Lambda$, $\widehat{\Lambda'}$ satisfy
\begin{eqnarray} 
\widehat{[\Lambda,{\Lambda'}]}=[\widehat\Lambda,\widehat{\Lambda'}] 
+ i \delta_\Lambda \widehat{\Lambda'}
- i \delta_{\Lambda'} \widehat\Lambda \,.
\label{8}
\end{eqnarray}

{\it Enveloping algebra-valued noncommutative gauge parameters and fields}, i.e.
for the enveloping algebra-valued gauge transformation, the commutator 
\begin{eqnarray}
[\widehat\Lambda,\widehat\Lambda'] 
=  \frac{1}{2}\{\Lambda_a(x)\stackrel{\star}{,} \Lambda'_b(x)\}[T^a,T^b] 
+ \frac{1}{2} [\Lambda_a(x)\stackrel{\star}{,}\Lambda'_b(x)]\{T^a,T^b\}
\label{9}
\end{eqnarray}
of two Lie algebra-valued noncommutative gauge parameters 
$\widehat\Lambda = \Lambda_a(x) T^a$ and $\widehat\Lambda' = \Lambda'_a(x) T^a$
does not close in Lie. For noncommutative SU(N) the Lie algebra 
traceless condition is incompatible with commutator. 
So, for noncommutative gauge transformation we have extension to 
the enveloping algebra-valued gauge transformation.

{\it Seiberg-Witten map:}\\
Closing condition for gauge transformation algebra are homogenus differential equations,
which are solved by iteration, order by order in noncommutative parameter $\theta$. 
Solutions are known as Seiberg-Witten map. Hermicity condition for the fields, 
up to the first order in Seiberg-Witten expansion, 
gives for gauge parameter, gauge and fermion fields
the following expressions:  
\begin{eqnarray} 
\widehat \Lambda &=& \Lambda +
\frac{h}{4} \theta^{\mu \nu}
           \{ V_{\nu}, \partial_{\mu}\Lambda\}+...
\nonumber\\
\widehat V_\mu  &=& V_{\mu} +
    \frac{h}{4} \theta^{\alpha \beta}
           \{ \partial_{\alpha}V_{\mu} + F_{\alpha \mu},V_{\beta}\}+...
\nonumber\\
\widehat\psi &=&  \psi - \, \frac{h}{2}\,\theta^{\alpha \beta} \,
\,\Big( V_{\alpha}  \,\partial_{\beta}
          - \frac{\mathrm{i}}{4}  \,
              [V_{\alpha}, V_{\beta}] \Big) \,\psi+...	      
\label{11}
\,.
\end{eqnarray}\\

{\it Extended noncommutative gauge field theory} 

Commutative GFT, that are renormalizable
 with minimal coupling, are extended 
in the same minimal fashion 
to the NC space with deformed gauge transformations. 
These deformations are not unique. For instance deformed action ${S_g}$
depends on the choice of representation. 
This derives from the fact that ${\widehat F}^{\mu\nu}$ is enveloping algebra
not Lee algebra valued. So called extended gauge-invariant action is:
\begin{eqnarray}
S_{\rm NC}&=& S_g+S_{\psi}
=-\frac{1}{2}\;{\rm Tr}\;\int d^4x\; 
\left(1-\frac{a-1}{2}h\theta_{\rho\sigma} \star \widehat F^{\rho\sigma} \right)
\star 
{\widehat F}_{\mu\nu} \star {\widehat F}^{\mu\nu}
\nonumber \\
&+&\mathrm{i}\;\int \mathrm{d}^4x\; \widehat{\bar \varphi}
\star \bar\sigma^\mu (\partial_\mu +\mathrm{i}\widehat A_\mu ) 
\star \widehat\varphi\,.
\label{12} 
\end{eqnarray}
The trace Tr in ${S_g}$ is over all representations.
${\widehat{\varphi}}$'s are the noncommutative Weyl spinors.
Applying Seiberg-Witten map on the above action up to first order in $\theta$
we obtain `minimal' actions
\begin{eqnarray}
S_g
&=& 
- \frac{1}{2} {\rm Tr} \int d^4 x \;
  F_{\mu \nu} F^{\mu \nu}
\nonumber \\ 
&+&h\,\theta^{\rho \sigma} {\rm Tr} \int d^4 x \;
\left[ \left( \frac{a}{4} F_{\rho \sigma} F_{\mu \nu}
-F_{\rho \mu} F_{\sigma \nu} \right) F^{\mu \nu} \right]\,,
\nonumber \\
S_{\psi}
&=&
\mathrm{i}\;\int d^4 x \;\bar\varphi\sigma^\mu (\partial_\mu
+\mathrm{i}  A_\mu )\varphi
\nonumber \\
&-&\frac{h}{8} \theta^{\mu\nu}
\Delta^{\alpha\beta\gamma}_{\mu\nu\rho}
\;\int d^4 x \;F_{\alpha\beta} \;\bar\varphi
\;\bar\sigma^\rho(\partial_\gamma
+\mathrm{i} A_\gamma)\varphi\,,
\nonumber\\
\Delta^{\alpha\beta\gamma}_{\mu\nu\rho}
&=& \varepsilon^{\alpha\beta\gamma\lambda}\varepsilon_{\lambda\mu\nu\rho}\,.
\label{13}
\end{eqnarray}

Clearly we do not know the meaning of `minimal coupling
concept' for some NCGFT in the NC space.  
However, renormalization is the principle that help us to find such 
acceptable couplings.
We learned that the renormalizability condition of some specific NCGFT 
requires introduction of the higher order noncommutative gauge interaction by
expanding general NC action in terms of NC field strengths.
This of course extends `NC minimal coupling' of 
the gauge action ${S_g}$ in (\ref{12}) to higher order;
with $a$ being free parameter determining renormalizable deformation.
This was possible due to the symmetry property of an object
$\theta_{\rho\sigma} \star \widehat F^{\rho\sigma}$.
SW map for NC field strength up to the first order in ${h\theta^{\mu\nu}}$ than gives:

In the chiral fermion sector the choice of Majorana spinors for the U(1) case gives
\begin{eqnarray}
S_{\psi}
&=&
\frac{\mathrm{i}}{2}\;\int d^4 x 
\Big[ \;\bar\psi\gamma^\mu (\partial_\mu
-\mathrm{i} \gamma_5 A_\mu )\psi
\nonumber \\
&+&\mathrm{i}\frac{h}{8} \theta^{\mu\nu}
\Delta^{\alpha\beta\gamma}_{\mu\nu\rho}
F_{\alpha\beta} \;\bar\psi
\;\gamma^\rho(\partial_\gamma
-\mathrm{i}\gamma_5 A_\gamma)\psi\,\Big]\,.
\label{16}
\end{eqnarray}
For the SU(2) case relevant expressions are given in \cite{Buric:2007ix}.

Proposed framework gives starting action
for the gauge and fermion sectors.
Requirement of renormalizability
fixes the freedom parameter $a$. That is, the
principle of renormalization determines NC renormalizable deformation.
Trace of three generators in the above action lead to  
dependence of the gauge group representation and
the choice of the trace corresponds to the choice of the group representation.
\\

{\it The gauge sector of minimal NCSM} \\
Here we choose vector field in the adjoint representation, i.e.  using a sum of three
traces over the standard model gauge group we obtain the following action
in terms of physical fields,
\begin{eqnarray}
S^{\mbox{\tiny mNCSM}}_g &=&
-\frac{1}{2} \int d^4x \Big[\left( \frac{1}{2} {\cal A}_{\mu\nu}{\cal A}^{\mu\nu}
+ \, \mbox{Tr}\, {\cal B}_{\mu\nu} {\cal B}^{\mu\nu}
+ \, \mbox{Tr}\, G_{\mu\nu} G^{\mu\nu} \right)
\nonumber
\\ & & 
- \frac{1}{2} \, g_s\, {d^{abc}} \, h\theta^{\rho\sigma}
\left(
\frac{ a}{4}G^a_{\rho\sigma} G^b_{\mu\nu}
-G^a_{\rho\mu} G^b_{\sigma\nu}
\right)G^{\mu\nu,c}\Big]  \, ,
\label{18}
\end{eqnarray}
where $d^{abc}$ are totally symmetric SU(3) group coefficients
which come from the trace in (\ref{13}). 
The ${\cal A}_{\mu\nu}$, ${\cal B}_{\mu\nu}(=B_{\mu\nu}^aT_L^a)$
and $G_{\mu\nu}(=G_{\mu\nu}^aT_S^a)$ denote the U(1), 
$\rm SU(2)_L$ and $\rm SU(3)_c$ field strengths, respectively.

For adjoint representation their is no new neutral electroweak triple gauge boson interactions.\\

{\it The nonminimal NCSM gauge sector action} \\
is obtained by
taking trace over all massive particle multiplets with different quantum numbers 
in the model that have covariant derivative acting on them;
five multiplets for each generation of fermions and one Higgs multiplet.
Here $V_{\mu}$ is the standard model gauge potential:
\begin{eqnarray} 
V^{\mu}&=&g'{\cal A}^{\mu}(x)Y + g\sum^3_{a=1}B^{\mu}_a(x)T^a_L 
+ g_s\sum^8_{b=1}G^{\mu}_{b}(x)T^b_S\,.
\label{20}
\end{eqnarray}
For detailes of the model see \cite{Goran}.

The interactions Lagrangian's in terms of physical fields and effective couplings are \cite{Goran}:
\begin{eqnarray}
{\cal L}^{\theta}_{\gamma\gamma\gamma}&=&\frac{e}{4} \sin2{\theta_W}\;
{{\rm K}_{\gamma\gamma\gamma}}
h\theta^{\rho\tau}A^{\mu\nu}\left({ a}A_{\mu\nu}A_{\rho\tau}-4A_{\mu\rho}A_{\nu\tau}\right)\,,
\label{22}\\
& & \nonumber \\
{\cal L}^{\theta}_{Z\gamma\gamma}&=&\frac{e}{4} \sin2{\theta_W}\,
{{\rm K}_{Z\gamma \gamma}}\,
h\theta^{\rho\tau}
\left[2Z^{\mu\nu}\left(2A_{\mu\rho}A_{\nu\tau}-{ a}A_{\mu\nu}A_{\rho\tau}\right)
\right.\nonumber\\
& & +\left. 
8 Z_{\mu\rho}A^{\mu\nu}A_{\nu\tau} - 
{ a}Z_{\rho\tau}A_{\mu\nu}A^{\mu\nu}\right]\,,\;\;\;{\rm ect.},
\label{23} \\
{\rm K}_{\gamma\gamma\gamma}&=&\frac{1}{2}\; gg'(\kappa_1 + 3 \kappa_2)\,,
\;\;\;\;\;
{\rm K}_{Z\gamma\gamma}=\frac{1}{2}\; 
\left[{g'}^2\kappa_1 + \left({g'}^2-2g^2\right)\kappa_2\right]\,,\;\;\;{\rm ect.} 
\nonumber
\end{eqnarray} \\

{\it Renormalization}

One-loop renormalization is performed by using 
the background field method (BFM).
Advantage of the BFM is the guarantee
of covariance, because by doing the path integral the local symmetry
of the quantum field ${\Phi}_V$ is fixed, while the gauge
symmetry of the background field ${\phi}_V$ is manifestly preserved.
Quantization is performed by the
functional integration over the quantum vector field ${\bf
\Phi}_V$ in the saddle-point approximation around classical
(background) configuration. For case $\phi _V= constant$,
the main contribution to the
functional integral is given by the Gaussian integral.
Split the vector potential
into the classical background plus the quantum-fluctuation parts,
that is: We replace, $\phi_V\to \phi_V + {\bf\Phi}_V$, and than compute the
terms quadratic in the quantum fields. 
Interactions are of the polynomial type.

Proper quantization requires the presence
of the gauge fixing term $ {S_{\rm gf}[\phi]}$. 
Adding to the SM part in the usual way,
Feynman-Faddeev-Popov ghost appears in the effective action. 
Result of functional integration 
\begin{eqnarray}
\Gamma [\phi] = S_{\rm cl}[\phi] + S_{\rm gf}[\phi] + \Gamma^{(1)}[\phi]\,,
\;\;\;
 S_{\rm gf}[\phi]=-\frac 12 \int
\mathrm{d}^4x(D_{\mu}\bf\Phi_V^{\mu})^2, 
\label{25}
\end{eqnarray}
produce the standard result of the commutative part of our action.
The one-loop effective part $\Gamma^{(1)} [\phi]$ is given by
\begin{equation}
\Gamma ^{(1)}[\phi] =\frac{\mathrm{i}}{2}\log\det
S^{(2)}[\phi]=\frac{\mathrm{i}}{2} {\rm Tr}\log S^{(2)}[\phi]\,,
\label{26}
\end{equation}
where $S^{(2)}[\phi]$ is the second functional derivative of a classical action.

The one-loop effective action computed by using background field method gives
 noncommutative vertices; see detailes in
\cite{Buric:2005xe,Buric:2006wm,Latas:2007eu,Buric:2007ix}.\\

{\it Renormalization of nmNCSM}

Divergences for $\rm U(1)_Y-SU(2)_C$ and $\rm U(1)_Y-SU(3)_C$ mixed 
noncommutative terms are
\begin{eqnarray}
\Gamma^{(1)}_{\rm div}&=& \frac {11}{3 (4\pi)^2\epsilon}\int d^4 x B_{\mu\nu}^iB^{\mu\nu i} +
\frac {11}{2 (4\pi)^2\epsilon}\int d^4 x G_{\mu\nu}^aG^{\mu\nu a}
\label{28}\\ 
&+& 
\frac{4}{3(4\pi)^2\epsilon}g^\prime g^2  \kappa_2(3-{ a})h\theta^{\mu\nu}\int d^4 x  
\big(\frac 14 f_{\mu\nu}B_{\rho\sigma }^i B^{\rho\sigma i} - f_{\mu\rho}B_{\nu\sigma }^i B^{\rho\sigma i}
 \big)
 \nonumber \\
&+& \frac{6}{3(4\pi)^2\epsilon}g^\prime g^2_S \kappa_3(3-{ a})h\theta^{\mu\nu}\int d^4 x  
\big(\frac 14 f_{\mu\nu}G_{\rho\sigma }^a G^{\rho\sigma a} - 
f_{\mu\rho}G_{\nu\sigma }^a G^{\rho\sigma a}
 \big)\,.
\nonumber  
 \end{eqnarray}
Renormalization is obtained via counter-terms and for the obvious choice $a=3$,
giving bare Lagrangian.
Constants $\kappa_1$, $\kappa_2$ and $\kappa_3$ remain  
unchanged under renormalization
if specific replacement in $1/g^2_i$ couplings were applied, see \cite{Buric:2006wm}.
Since, for $a=3$, our Lagrangian is free from divergences at one-loop
noncommutative deformation parameter $h$ need not be renormalized.\\

{\it Renormalization of NC SU(N) gauge theory}

Choosing vector field in the adjoint representation SU(N) we obtain
the starting Lagrangian (\ref{18}), 
where now $d^{abc}$ are totally symmetric SU(N) group coefficients.

Renormalization of the theory is obtained by canceling  divergences.
To have that the counter terms should be
added to the starting action, which than produces the bare Lagrangian
which has two solutions: $a=1$ and $a=3$ \cite{Latas:2007eu}.

The case $a=1$ corresponds to previous result \cite{Buric:2005xe} and  
the deformation parameter ${h}$ 
need not to be renormalized. Renormalizability is, 
in this case, obtained through the known renormalization 
of gauge fields and coupling constant only.

However the case $a=3$ is different since additional divergences can be absorbed
only into the noncommutative deformation parameter ${h}$. That is that ${h}$ has
to be renormalized. The bare gauge field, the coupling constant and 
the noncommutative deformation parameter are \cite{Latas:2007eu}:
\begin{eqnarray}
V^\mu_0=V^\mu\sqrt{1+\frac{22Ng^2}{3(4\pi)^2
\epsilon}},
\;\;\;
g_0=\frac{g\mu^{\epsilon/2}}{\sqrt{1+\frac{22Ng^2}{3(4\pi)^2
\epsilon}}},
\;\;\;
h_0=\frac{h}{1-\frac{2Ng^2}{3(4\pi)^2\epsilon}}\,. 
\label{36}
\end{eqnarray}
The necessity of the $ {h}$ renormalization jeopardizes previous hope
that the {NC SU(N)} gauge theory might be renormalizable to all orders
in $ {\theta^{\mu\nu}}$. Above results are also valid for 
the minimal NCSM gauge sector (\ref{18}) with $N=3$. \\

{\it Ultraviolet asymptotic behavior of NC SU(N) gauge theory}

Gauge coupling constant ${g}$ and the NC deformation
parameter ${h}$ in our theory 
depend on energy i.e., the renormalization point ${\mu}$.
Both beta functions ($\beta_g,\;\beta_h$) are {\it negative} that is they
decrease with increasing energy ${\mu}$ \cite{Latas:2007eu}. 
Solution to ${\beta_h}$ shows that by increase of energy ${\mu}$ the NC deformation
parameter ${h}$ decreases \cite{Latas:2007eu}. From this follows necessity of the 
modification of Heisenberg uncertainty relations at high energy. 
String theory inspired modification
\begin{eqnarray}
[x,p]=i\hbar(1+\beta p^2)
\;\;\;\;\Rightarrow\;\;\;\;
\Delta x=\frac{\hbar}{2}(\frac{1}{\Delta p}+\beta \Delta p).
\label{39}
\end{eqnarray}
show that for large momenta ${\Delta p}$ (energy) distance ${\Delta x}$ grows linearly.  
So large energies do not necessarily correspond to small distances, and running ${h}$ does
not imply that noncommutativity vanishes at small distances.
This is related to {UV/IR} correspondence. From Eq. (\ref{36}) and $h=1/\Lambda_{\rm NC}$
we have
\begin{equation}
h(\mu)=\frac{1}{\Lambda^2_{\rm NC}(\mu)}
\;\;\;\;\Rightarrow\;\;\;\;
\Lambda_{\rm NC}(\mu) =\Lambda_{\rm NC}\,\sqrt{\ln
\frac{\mu}{\Lambda}}\,. 
\label{40}
\end{equation}
This way, via RGE, the scale of noncommutativity ${\Lambda_{\rm NC}}$ 
becomes the running scale of noncommutativity \cite{Latas:2007eu}. 
However it receives very small change when energy ${\mu}$ increases.\\

{\it The 4$\psi$ divergences for NC chiral fermions in U(1) and SU(2) cases}

The one-loop effective action is computed from Eq. (\ref{16}) by using 
background field method.

Our computations shows that divergent contributions from relevant terms  
are finite due to the structure of the momentum integrals in both, the U(1)
and the SU(2), cases.
For NC chiral electrodynamics, with Majorana spinors 
and with the usual definition for the supertrace $\mathrm{STr}$, \cite{Buric:2004ms},
we have used chiral fermions in the fundamental representation of SU(2). 
Choosing Majorana spinors we apparently break the SU(2) symmetry, and consequently  
we have to work in the framework of the components for the vector potential.
Now of course, Majorana ${\psi_1 \choose \psi_2} $ is not a SU(2) doublet.

Clearly, we conclude that direct computations by using BFM confirms results of 
the symmetry analysis for the 4$\psi$ divergent terms, 
which, due to their U(1) invariance, have to be zero \cite{Buric:2007ix}.
The same symmetry arguments holds also SU(2) terms,
i.e. they both vanish identically too.\\

{\it Limits on the noncommutativity scale}

From  the gauge-invariant amplitude for $Z \to \gamma\gamma,\;gg$ decays in momentum space
and for $Z$ boson at rest and for $a=3$, we have found the following branching ratios
\cite{Buric:2007qx},
\begin{eqnarray}
BR_{(Z\rightarrow \gamma\gamma)}
=\tau_Z \frac{\alpha}{4}\frac{M^5_Z}{\Lambda^4_{\rm NC}}\sin^2 2\theta_W
{\rm K}^2_{Z\gamma \gamma}
\big({\vec E}_{\theta}^2 + {\vec B}_{\theta}^2\big)
=\frac{1}{8}\frac{{\rm K}^2_{Z\gamma \gamma}}{{\rm K}^2_{Zgg}}BR_{(Z\rightarrow gg)},
\label{BR}
\end{eqnarray}
where $\tau_Z $ is the $Z$ boson lifetime. 
LHC experimental possibilities for $Z \to \gamma\gamma$ 
we analyze by using the CMS Physics Technical Design Report \cite{CMS1,CMS2}.
We have found that for $10^7$ events of $Z\rightarrow e^+e^-$ for $10\;fb^{-1}$ in 2 years of LHC
running and by assuming $BR(Z\rightarrow \gamma\gamma) \sim 10^{-8}$ and 
using $BR(Z\rightarrow e^+e^-) = 0.03 $ about $ \sim 3$ events
of $Z\rightarrow \gamma\gamma$ decays should be found. 
However, note that background sources (CMS Note 2006/112, Fig.3) could potentially be a big problem.
For example study for $Higgs\rightarrow \gamma\gamma$ shows that, 
when ${e^-}$ from $Z\rightarrow e^+e^-$  radiates 
very high energy Bremsstrahlung photon
into pixel detector, for similar energies of 
${e^-}$ and ${\gamma}$, there is a
huge probability of misidentification of 
${e^-}$ with ${\gamma}$. Second, the irreducible di-photon background
may also kill the signal. The $Z \to gg$ decay was discussed in \cite{Goran}.

Finally, note that after 10 years of LHC running integrated 
luminosity would reach $\sim  1000\;fb^{-1}$.
In that case and from bona fide reasonable assumption $BR(Z\rightarrow \gamma\gamma) \sim 10^{-8}$ 
one would find ~$300$ events of $Z\rightarrow \gamma\gamma$
decays, or one would have $ \sim 3$ events with 
$BR(Z\rightarrow \gamma\gamma) \sim 10^{-10}$.
From above it follows that, in the later case, the lower bound on the
scale of noncommutativity would be $\Lambda_{\rm NC}\,>\, 1\; {\rm TeV}$.\\


Limits on the scale of noncommutativity in high energy particle physics 
are coming from the analysis of decay and scattering experiments. 

Considering SM forbidden decays, recently we have found 
the following lower limit $\Lambda_{\rm NC} > 1$ TeV \cite{Buric:2007qx}
from $Z\rightarrow \gamma\gamma$ decay.
Note here that earlier limits obtained from 
 $\gamma_{\rm pl}\to \nu \bar\nu$ decay (astrophysics analysis) produces
$\Lambda_{\rm NC} > 81$ GeV \cite{Josip}
while from the SM forbidden
$J/\psi\to\gamma\gamma$ and $K\to\pi\gamma$ 
\cite{Melic:2005hb} 
decays we obtain  
$\Lambda_{\rm NC} > 9$ GeV, and
$\Lambda_{\rm NC} > 43$ GeV, respectively. 
Last two bounds are not usefull 
due to the too high lower limit of the relevant branching ratios.

Scattering experiments \cite{Ohl:2004tn} support the above obtained limits.
From annihilation $\gamma\gamma\to\,{\bar f}f$ it was found $\Lambda_{\rm NC} > 200$ GeV,
which is a bit to low. However, 
from ${\bar f}f \to Z\gamma $ unelastic scattering experiments
there is very interesting limit $\Lambda_{\rm NC} > 1$ TeV.\\

{\it Summary and Conclusion}

Principle of renormalizability implemented on 
our {$\theta$}-expanded NCGFT led us to well defined deformation via introduction of
higher order noncommutative action class for the gauge sectors of 
the mNCSM, nmNCSM and NC SU(N) models.
This extension was parametrized by generically free parameter $a$:
\begin{equation}
 S_g =-\frac{1}{2}{\rm Tr}\int d^4x
\left(1
+\mathrm{i}({a}-1)\,\widehat x_{\rho}\star \widehat x_{\sigma}
\star\widehat F^{\rho\sigma}\right)\star\widehat F_{\mu\nu}
\star\widehat F^{\mu\nu}\,.
\label{46}
\end{equation}
\normalsize
We have found the following properties of the above models 
with respect to renormalization procedure:
\\
- Renormalization principle is fixing the freedom parameter $a$
for our $ \theta$-expanded NC GFT.
\\
- Divergences cancel differently than in 
commutative GFT and this depends on the representations.
\\
- Gauge sector of the nmNCSM, which produces SM forbidden $Z\to \gamma\gamma$ decay, is
renormalizable and {\it finite} for ${a=3}$. 
Due to this finiteness no renormalization of ${h}$ necessary.
\\
- Noncommutative SU(N) gauge theory is renormalizable for ${a=1}$ and ${a=3}$. 
The case ${a=1}$ corresponds to the earlier obtained result \cite{Buric:2005xe}. 
However, in the case ${a=3}$
additional divergences appears and had to be absorbed through the renormalization
of the noncommutative deformation ${h}$. Hence, in the case of noncommutative SU(N) 
the noncommutativity deformation parameter 
$ {h}$  had to be renormalized and it is 
{\it asymptotically free}, opposite to the previous expectations. 
The same is valid for mNCSM. 
\\
- The solution ${a=3}$, while shifting the model
to the higher order, i.e. while extending `NC minimal coupling', 
hints into the discovery of  
the key role of the higher noncommutative gauge interaction in
one-loop renormalizability of classes of NCGFT at the first order in $\theta$. 
\\
- Our computations also confirms symmetry arguments that 
for noncommutative chiral electrodynamics, 
that is the U(1) case with Majorana spinors, 
the $4\psi$ divergent part vanishes.
For noncommutative chiral fermions in the fundamental representation of SU(2)
with Majorana spinors the $4\psi$ divergent part vanishes due to the SU(2) invariance.
So, for noncommutative U(1) and SU(2) chiral fermion models 
typical $4\psi$-divergence is {\it absent}, contrary to the earlier results obtained  
for Dirac fermions \cite{Wulkenhaar:2001sq,Buric:2004ms}.
\\
- There is similarity to noncommutative $\phi^4$ theory \cite{Grosse:2005da}. 
\\
- Note that the
renormalizability principle could help to minimize or even cancel
most of the ambiguities of the higher order Seiberg-Witten
maps~\cite{Moller:2004qq}.
\\
- Finally, phenomenological results, as the standard model forbidden
$Z\to \gamma\gamma$ decay, are 
{\it robust} due to the one-loop renormalizability and
{\it finiteness} of the nmNCSM gauge sector \cite{Buric:2006wm,Buric:2007qx}.

\section*{Acknowledgment}
Part of this work was done during my visit to ESI, Vienna and MPI, M\" unchen.
I would like to use this opportunity to acknowledge 
H. Grosse at ESI, and W. Hollik at MPI, for hospitality and support.


%


\begin{thebibliography}{}
 \bibitem{Wess}
M.~Kontsevich,
{\it Deformation quantization of Poisson manifolds, I},
Lett.\ Math.\ Phys.\  {\bf 66} (2003) 157
[q-alg/9709040].
%
  %
 \bibitem{Seiberg:1999vs}
N. Seiberg and E. Witten, 
{\it String theory and noncommutative geometry}, 
JHEP {\bf 09} (1999) 032 [arXiv:hep-th/9908142].
%
J. Madore, S. Schraml, P.~Schupp and J.~Wess, 
{\it Gauge theory on noncommutative spaces}, 
Eur. Phys. J. C{\bf 16} (2000) 161 [arXiv:hep-th/0001203];
%
B. Jur\v{c}o, S. Schraml, P.~Schupp and J.~Wess, 
{\it Enveloping algebra valued gauge transformations for non-Abelian gauge groups on
non-commutative spaces}, 
Eur. Phys. J. C{\bf 17} (2000) 521 [arXiv:hep-th/0006246];
%
B. Jur\v{c}o, L. M\"oller, S.~Schraml, P.~Schupp and J.~Wess, 
{\it Construction of non-Abelian gauge theories on non-commutative spaces}, 
Eur. Phys. J. C{\bf 21} (2001) 383 [arXiv:hep-th/0104153].
 %
\bibitem{Calmet:2001na}
X.~Calmet, B.~Jur\v{c}o, P.~Schupp, J.~Wess and M.~Wohlgenannt, 
{\it The standard model on noncommutative space-time}, 
Eur.~Phys.~J. C{\bf 23} (2002) 363 [arXiv:hep-ph/0111115].
%
\bibitem{Blazenka}
B.~Melic, K.~Passek-Kumericki, J.~Trampetic, P.~Schupp and M.~Wohlgenannt,
{\it The standard model on noncommutative space-time: Electroweak currents  and Higgs sector},
  Eur.\ Phys.\ J.\ C {\bf 42} (2005) 483 [arXiv:hep-ph/0502249];
{\it The standard model on noncommutative space-time: Strong interactions included},
  ibid 499 [arXiv:hep-ph/0503064].
%
\bibitem{Aschieri:2002mc}
P. Aschieri, B.~Jur\v{c}o, P.~Schupp and J.~Wess, 
{\it Noncommutative GUTs, Standard Model and C,P,T}, 
Nucl. Phys. B{\bf 651} (2003) 45 [arXiv:hep-th/0205214].
%
\bibitem{Goran}
W.~Behr, N.G. Deshpande, G. ~Duplan\v{c}i\'{c}, P.~Schupp, J.~Trampeti\'{c} and J.~Wess, 
{\it The Z $\to\,\,\gamma\gamma,gg$ Decays in the Noncommutative Standard Model}, 
Eur. Phys. J. C{\bf 29} (2003) 441 [arXiv:hep-ph/0202121];
%
G.~Duplan\v{c}i\'{c}, P.~Schupp and J.~Trampeti\'{c}, 
{\it Comment on triple gauge boson interactions in the noncommutative electroweak sector}, 
Eur.~Phys. J. C{\bf 32} (2003) 141 [arXiv:hep-ph/0309138].
%
\bibitem{UV/IR}
%
  S.~Minwalla, M.~Van Raamsdonk and N.~Seiberg,
{\it Noncommutative perturbative dynamics},
  JHEP {\bf 0002} (2000) 020
  [arXiv:hep-th/9912072];
  %
\bibitem{Seiberg:2000gc}
  N.~Seiberg, L.~Susskind and N.~Toumbas,
  {\it Space/time noncommutativity and causality},
  JHEP {\bf 0006}, 044 (2000)
  [arXiv:hep-th/0005015].
\bibitem{Gomis:2000zz}
  J.~Gomis and T.~Mehen,
  {\it Space-time noncommutative field theories and unitarity},
  Nucl.\ Phys.\ B {\bf 591}, 265 (2000) [arXiv:hep-th/0005129].
\bibitem{Carroll:2001ws}
 S.~M.~Carroll, J.~A.~Harvey, V.~A.~Kostelecky, C.~D.~Lane and T.~Okamoto,
 {\it Noncommutative field theory and Lorentz violation},
  Phys.\ Rev.\ Lett.\  {\bf 87}, 141601 (2001) [arXiv:hep-th/0105082].
%
\bibitem{Brandt:2003fx}
F. Brandt, C.P. Martin and F. Ruiz Ruiz, 
{\it Anomaly freedom in Seiberg-Witten noncommutative gauge theories}, 
JHEP {\bf 07} (2003) 068 [arXiv:hep-th/0307292].
%
\bibitem{Bichl:2001cq}
A.~Bichl, J. Grimstrup, H. Grosse, L. Popp, M. Schweda and R. Wulkenhaar, 
{\it Renormalization of the noncommutative photon
self-energy to all orders via Seiberg-Witten map}, 
JHEP {\bf 06} (2001) 013 [arXiv:hep-th/0104097].
%
\bibitem{Buric:2005xe}
  M.~Buric, D.~Latas and V.~Radovanovic,
  {\it Renormalizability of noncommutative SU(N) gauge theory},
  JHEP {\bf 0602} (2006) 046 [arXiv:hep-th/0510133]\,;
\bibitem{Buric:2006wm}
  M.~Buric, V.~Radovanovic and J.~Trampetic,
{\it The one-loop renormalization of the gauge sector in 
the $\theta$-expanded noncommutative standard model},
  JHEP {\bf 03} (2007) 030 [arXiv:hep-th/0609073].
  %
 %
\bibitem{Latas:2007eu}
  D.~Latas, V.~Radovanovic and J.~Trampetic,
 {\it Noncommutative SU(N) gauge theories and asymptotic freedom},
  Phys.\ Rev.\  D {\bf 76} (2007) 085006, arXiv:hep-th/0703018.
  %
\bibitem{Martin:2006gw}
  C.~P.~Martin, D.~Sanchez-Ruiz and C.~Tamarit,
  {\it The noncommutative U(1) Higgs-Kibble model in the enveloping-algebra
  formalism and its renormalizability},
JHEP {\bf 0702} (2007) 065
  [arXiv:hep-th/0612188].
  %
  %
\bibitem{Buric:2007ix}
  M.~Buric, D.~Latas, V.~Radovanovic and J.~Trampetic,
 {\it The absence of the 4$\psi$ divergence in noncommutative chiral models},
 Phys.\ Rev.\  D {\bf 77} (2008) 045031; arXiv:0711.0887 [hep-th].
  %
\bibitem{Buric:2007qx}
  M.~Buric, D.~Latas, V.~Radovanovic and J.~Trampetic,
 {\it Nonzero Z $\to$ gamma gamma decays in the renormalizable gauge sector of
  the noncommutative standard model},
  Phys.\ Rev.\  D {\bf 75} (2007) 097701 [arXiv:hep-ph/0611299].
%
  \bibitem{Josip}
%
  P.~Schupp, J.~Trampetic, J.~Wess and G.~Raffelt,
 {\it The photon neutrino interaction in noncommutative gauge field theory  and
  astrophysical bounds},
  Eur.\ Phys.\ J.\  C {\bf 36} (2004) 405
  [arXiv:hep-ph/0212292].
  P.~Minkowski, P.~Schupp and J.~Trampetic,
 {\it Neutrino dipole moments and charge radii in noncommutative space-time},
  Eur.\ Phys.\ J.\  C {\bf 37} (2004) 123
  [arXiv:hep-th/0302175]\,.
%
\bibitem{Melic:2005hb}
  B.~Melic, K.~Passek-Kumericki and J.~Trampetic,
{\it Quarkonia decays into two photons induced by the space-time noncommutativity},
  Phys.\ Rev.\ D {\bf 72} (2005) 054004 [arXiv:hep-ph/0503133];
  {\it $K \to \pi \gamma$ decay and space-time noncommutativity},
  ibid 057502 [arXiv:hep-ph/0507231];
  C.~Tamarit and J.~Trampetic,
  {\it Noncommutative fermions and quarkonia decays},
  arXiv:0812.1731 [hep-th].
%
%
\bibitem{CMS1} CMS Physics Technical Design Report, Vol.1.  CERN/LHCC 2006-001.
%
\bibitem{CMS2} M. Pieri et al., CMS Note 2006/112.
%
  %
   %
  \bibitem{Ohl:2004tn}
  A.~Alboteanu, T.~Ohl and R.~Ruckl,
  {\it Probing the noncommutative standard model at hadron collider},
  Phys.\ Rev.\ D {\bf 74}, 096004 (2006) [arXiv:hep-ph/0608155].
\bibitem{Wulkenhaar:2001sq}
  R.~Wulkenhaar,
  {\it Non-Renormalizability Of Theta-Expanded Noncommutative QED},
  JHEP {\bf 0203} (2002) 024 [arXiv:hep-th/0112248].
%
\bibitem{Buric:2004ms}
  M.~Buric and V.~Radovanovic,
  {\it Non-renormalizability of noncommutative SU(2) gauge theory},
  JHEP {\bf 0402} (2004) 040 [arXiv:hep-th/0401103];
  %
    %
\bibitem{Grosse:2005da}
  H.~Grosse and R.~Wulkenhaar,
 {\it Renormalization of $\phi^4$-theory on noncommutative $R^4$ to all orders},
  Lett.\ Math.\ Phys.\  {\bf 71}, 13 (2005);
 {\it Regularization and renormalization of quantum field theories on noncommutative spaces},
  J.\ Nonlin.\ Math.\ Phys.\  {\bf 11S1}, 9 (2004);
   H.~Grosse and H.~Steinacker,
 {\it Renormalization of the noncommutative $\phi^3$ model through the  Kontsevich model},
  Nucl.\ Phys.\  B {\bf 746}, 202 (2006)
  [arXiv:hep-th/0512203];
  %
   H.~Grosse and M.~Wohlgenannt,
 {\it Renormalization and Induced Gauge Action on a Noncommutative Space},
    Prog.\ Theor.\ Phys.\ Suppl.\  {\bf 171} (2007) 161
  [arXiv:0706.2167 [hep-th]].
  %
\bibitem{Moller:2004qq}
L.~M{\"o}ller,
{\it Second order of the expansions of action functionals of the
 noncommutative standard model},
  {\em JHEP} {\bf 10} (2004) 063 [arXiv:hep-th/0409085];
A.~Alboteanu, T.~Ohl and R.~R\"uckl,
{\it The Noncommutative Standard Model at $\mathcal O(\theta^2)$},
Phys.\ Rev.\  D {\bf 76} (2007) 105018, 0707.3595[hep-th];
Josip Trampeti\'{c} and Michael Wohlgenannt,
{\it Remarks on the second-order Seiberg-Witten maps},
Phys.\ Rev.\  D {\bf 76}, 127703 (2007), 0710.2182[hep-th].


\end{thebibliography}
\end{document}